\documentclass[reprint,twocolumn,floats,aps,english,preprintnumbers,amsmath,amssymb,array,prl]{revtex4-1}
\usepackage[T1]{fontenc}
\usepackage[latin9]{inputenc}
\usepackage{verbatim}
\usepackage{float}
\usepackage{graphicx}
\usepackage{prettyref}

\makeatletter
\@ifundefined{textcolor}{}
{%
 \definecolor{BLACK}{gray}{0}
 \definecolor{WHITE}{gray}{1}
 \definecolor{RED}{rgb}{1,0,0}
 \definecolor{GREEN}{rgb}{0,1,0}
 \definecolor{BLUE}{rgb}{0,0,1}
 \definecolor{CYAN}{cmyk}{1,0,0,0}
 \definecolor{MAGENTA}{cmyk}{0,1,0,0}
 \definecolor{YELLOW}{cmyk}{0,0,1,0}
 }

\usepackage{dcolumn}
\usepackage{bm}
\usepackage{enumerate}

\makeatother

\usepackage{babel}

\begin{document}

\title{Memory Effects In Nonequilibrium Quantum Impurity Models}

\author{Guy Cohen}
\affiliation{School of Chemistry, The Sackler Faculty of Exact Sciences,
  Tel Aviv University, Tel Aviv 69978, Israel}

\author{Eran Rabani}\
\affiliation{School of Chemistry, The Sackler Faculty of Exact Sciences,
  Tel Aviv University, Tel Aviv 69978, Israel}

\date{\today}

\begin{abstract}
Memory effects play a key role in the dynamics of strongly correlated
systems driven out of equilibrium. In the present study, we explore
the nature of memory in the nonequilibrium Anderson impurity model.
The Nakajima--Zwanzig--Mori formalism is used to derive an exact
generalized quantum master equation for the reduced density matrix of
the interacting quantum dot, which includes a non-Markovian memory
kernel. A real-time path integral formulation is developed, in which
all diagrams are stochastically sampled in order to numerically
evaluate the memory kernel. We explore the effects of temperature down
to the Kondo regime, as well as the role of source--drain bias voltage
and band width on the memory. Typically, the memory decays on
timescales significantly shorter than the dynamics of the reduced
density matrix itself, yet under certain conditions it develops a
smaller long tail. In addition we address the conditions required for
the existence, uniqueness and stability of a steady-state.
\end{abstract}
\maketitle

Interest in the problem of intrinsically \emph{nonequilibrium} open
quantum systems---in which one considers a small, strongly interacting
and highly correlated region coupled to several large, noninteracting
baths---has been surging in both experiment and theory. The aim of
theory in this regard is to provide a solid framework for
understanding phenomena ranging from the nonequilibrium Kondo effect
in quantum dots to conductance through single
molecules~\cite{nitzan_electron_2003}.  While much progress has been
made recently, based on brute-force approaches such as time-dependent
numerical renormalization group
techniques~\cite{anders_real-time_2005,white_density_1992,schmitteckert_nonequilibrium_2004}
and iterative~\cite{weiss_iterative_2008,segal_numerically_2010} or
stochastic~\cite{muehlbacher_real-time_2008,werner_diagrammatic_2009,schiro_real-time_2009}
diagrammatic approaches to real-time path integral formulations, the
problem has never been fully solved in a satisfactory manner. In fact,
it is becoming clearer that major gaps exist in our understanding of
the dynamics, the crossover regimes, the dependence on initial
conditions and the behavior at steady state.

The kind of open systems discussed above are often addressed by
impurity models, which explicitly account for the two types of regions
within the problem by partitioning the  Hamiltonian into system and
bath subspaces: $H=H_{S}+H_{B}+V$, where $H_{S}$ represents a
low-dimensional but interacting ``system'' subspace, $H_{B}$
represents a set of non-interacting lead or bath subspaces, and $V$ is
a system--bath coupling term. The dynamics generated by such
Hamiltonians can feature transients on timescales that are much longer
than the typical inverse energy scale~\cite{nordlander_how_1999},
where numerically exact approaches become intractable due to the
exponential growth of the active space or the equivalent complications
resulting from the dynamical sign problem. In many important
situations, however, the non-interacting baths can be traced out,
leading to a reduced description of the dynamics of the interacting
system at the cost of introducing non-locality in the time
propagation~\cite{feynman_theory_1963}. In path integral approaches
the effects of the leads are accounted for by a time non-local
influence
functional~\cite{muehlbacher_real-time_2008,weiss_iterative_2008}.

Perhaps a more appealing approach, which has been used to derive very
successful perturbative schemes for fermionic
systems~\cite{leijnse_kinetic_2008,li_coherent_2010} but is
notoriously difficult to carry out exactly, is based on the
generalized quantum master equation (GQME). In this formalism, a
so-called memory kernel replaces the influence functional. The
complexity of solving the many-body quantum Liouville equation is then
reduced to the evaluation of this memory kernel, which fully
determines the dynamics of the system. Furthermore, the memory kernel
contains all the information needed to resolve questions concerning
the existence and uniqueness of a
steady-state~\cite{dhar_nonequilibrium_2006,khosravi_bound_2009}, as
well as the values of system observables at steady state. While the
dynamical timescale of the system typically exceeds the characteristic
inverse energy scale, the memory kernel is expected to decay on
relatively short timescales for a large and interesting class of
physical situations (essentially whenever the bandwidth of the baths
is much larger than other energy scales in the problem). Thus,
brute-force approaches limited to short times are well suited to its
calculation.  Once the memory function is known, the formalism is
exact and tractable at \emph{all} times.

In this Letter, we explore the nature of memory in nonequilibrium
impurity models, and focus on the Anderson
problem~\cite{anderson_localized_1961}, covering the effects of
temperature down to the Kondo regime, the role of source--drain bias
voltages and band width. This is accomplished by adopting the
Nakajima--Zwanzig--Mori~\cite{nakajima_quantum_1958,zwanzig_ensemble_1960,mori_transport_1965}
formalism to derive an \emph{exact} GQME for the reduced density
matrix $\sigma\left(t\right)=P\rho\left(t\right)$ of the interacting
system, which includes a non-Markovian memory kernel. The conjecture
that the memory decays on timescales significantly shorter than the
dynamics of $\sigma\left(t\right)$ is confirmed, yet it is found that
it develops a smaller long tail when the Hubbard term is switched on.
The approach provides means to simulate the dynamics of the strongly
correlated subsystem on timescales beyond the limits of the path
integral method itself and reveals the conditions required for the
existence, uniqueness and stability of a steady-state under a finite
source-drain bias.

\begin{figure}
\includegraphics[width=1\columnwidth]{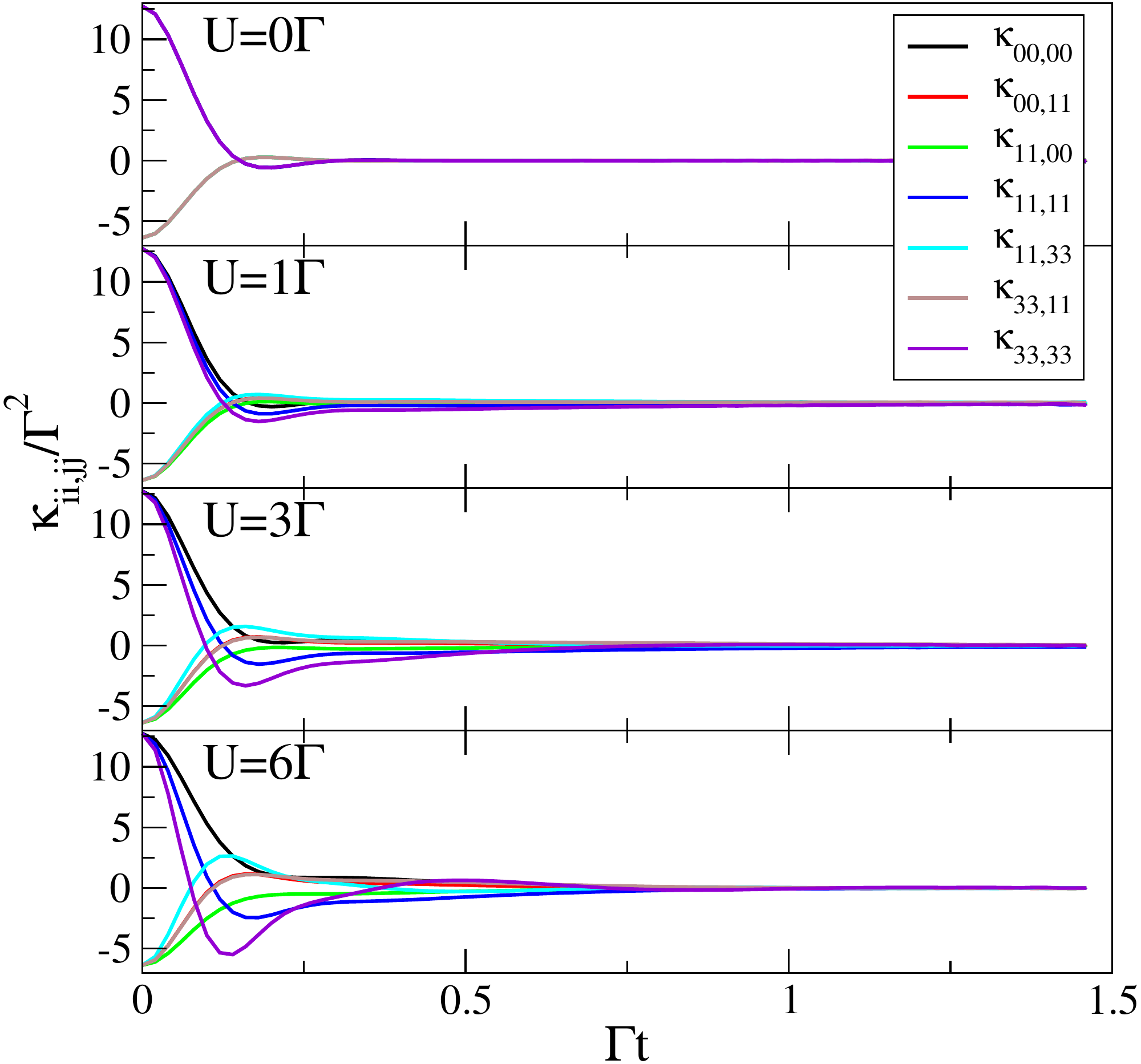}
\caption{Distinct nonzero memory kernel elements for an initially unoccupied
  dot and several values of the interaction energy $U$. The remaining
  parameters are
  $\frac{\mu_{L}}{\Gamma}=-\frac{\mu_{R}}{\Gamma}=\frac{1}{2}$,
  $\frac{\varepsilon_{c}}{\Gamma}=20$, and $\beta\Gamma=1$.
\label{fig:memory-kernel}}
\end{figure}

The exact equation of motion of the complete density matrix,
$i\hbar\frac{\mathrm{d}}{\mathrm{d}t}\rho=\left[H,\rho\right]$, is
governed by the Liouvillian $\mathcal{L}=\left[H,\,...\right]$.  If
the coupling term $V$ were to be turned off, the dynamics of the two
subspaces would be given by similar equations with the Liouvillian
replaced by the system and bath Liouvillians,
$\mathcal{L}_{S}=\left[H_{S},\,...\right]$ and
$\mathcal{L}_{B}=\left[H_{B},\,...\right]$, respectively. We also
define a coupling Liouvillian $\mathcal{L}_{V}=\left[V,\,...\right]$.
The equation of motion for the reduced density operator
$\rho_{B}\sigma=P\rho$ (using the projection operator onto the system
subspace $P=\rho_{B}\mathrm{Tr}_{B}\left\{ ...\right\} $) is then
given by:
\begin{equation}
i\hbar\frac{\mathrm{d}}{\mathrm{d}t}\sigma\left(t\right)=\mathcal{L}_{S}\sigma\left(t\right)+\vartheta\left(t\right)-\frac{i}{\hbar}\int_{0}^{t}\mathrm{d}\tau\,\kappa\left(\tau\right)\sigma\left(t-\tau\right),\label{eq:sigma_EOM}
\end{equation}
where $\vartheta\left(t\right)=\mathrm{Tr}_{B}\left\{
\mathcal{L}_{V}e^{-\frac{i}{\hbar}Q\mathcal{L}t}Q\rho_{0}\right\} $
contains the initial correlations,
$\kappa\left(t\right)=\mathrm{Tr}_{B}\left\{
\mathcal{L}_{V}e^{-iQ\mathcal{L}\tau}Q\mathcal{L}\rho_{B}\right\} $ is
the memory kernel in superoperator form, and $Q=1-P$. The initial
condition are contained within the initial density matrix $\rho_{0}$,
which also determines the initial bath part of the density matrix
$\rho_{B}=Q\rho_{0}$. 

The above equation for $\sigma\left(t\right)$ is exact, yet
requires as input the two superoperator functions $\kappa\left(t\right)$ and
$\vartheta\left(t\right)$, both of which have been defined in terms of projected propagation.
Evaluating such projected dynamics is cumbersome, and can be reduced to
solving a superoperator Volterra equation of the second type involving
a quantity $\Phi\left(t\right)$ which is free of projected
propagation~\cite{zwanzig_nonequilibrium_2001,zhang_nonequilibrium_2006}. For
the memory kernel one finds:
\begin{eqnarray}
\kappa\left(\tau\right) & = & i\hbar\dot{\Phi}\left(\tau\right)-\Phi\left(\tau\right)\mathcal{L}_{S}\nonumber \\
 &  & \,+\frac{i}{\hbar}\int_{0}^{\tau}\mathrm{d}\tau\Phi\left(t-\tau\right)\kappa\left(\tau\right),\label{eq:kappa_EOM}\\
\Phi\left(t\right) & = & \mathrm{Tr}_{B}\left\{ \mathcal{L}_{V}e^{-\frac{i}{\hbar}\mathcal{L}t}\rho_{B}\right\} .
\label{eq:Phi_definition}
\end{eqnarray}
A similar procedure exists for the initial correlation term
$\vartheta$; however, here we consider only the initially factorized
case $\rho_{0}=\rho_{B}\otimes\sigma\left(0\right)$, for which
$\vartheta=0$. As discussed below, the matrix elements of the
superoperator $\Phi$ are identical to quantities to which RT-PIMC has
already been
applied~\cite{muehlbacher_real-time_2008,werner_diagrammatic_2009}.

In the Anderson impurity model
$H_{S}=\sum_{i=\uparrow\downarrow}\varepsilon_{i}d_{i}^{\dagger}d_{i}+Ud_{\uparrow}^{\dagger}d_{\uparrow}d_{\downarrow}^{\dagger}d_{\downarrow}$,
$H_{B}=\sum_{k,i=\uparrow\downarrow}\varepsilon_{ik}a_{ik}^{\dagger}a_{ik}$
and
$V=\sum_{ki=\uparrow\downarrow}t_{ik}d_{i}a_{ik}^{\dagger}+t_{ik}^{*}a_{ik}d_{i}^{\dagger}$.
Thus, the system subspace is four-dimensional, being spanned by the
states $0\equiv\left|00\right\rangle \equiv\left|0\right\rangle
,\,1\equiv\left|01\right\rangle =d_{2}^{\dagger}\left|0\right\rangle
,\,2\equiv\left|10\right\rangle =d_{1}^{\dagger}\left|0\right\rangle $
and $3\equiv\left|11\right\rangle
=d_{1}^{\dagger}d_{2}^{\dagger}\left|0\right\rangle $.  With this
notation, we can perform a calculation to derive an expression for the
system Liouvillian
\begin{eqnarray}
\left[\mathcal{L}_{s}\right]_{ij,qq^{\prime}} & = & \delta_{iq}\delta_{jq^{\prime}}\left\{ \varepsilon_{1}\left(\delta_{q2}+\delta_{q3}-\delta_{j2}-\delta_{j3}\right)\right.\nonumber \\
 &  & \,+\varepsilon_{2}\left(\delta_{q1}+\delta_{q3}-\delta_{j1}-\delta_{j3}\right)\nonumber \\
 &  & \,\left.+U\left(\delta_{q3}-\delta_{j3}\right)\right\} ,
\end{eqnarray}
and for $\Phi$, which takes the more complicated form
\begin{eqnarray}
\Phi_{ij,qq^{\prime}} & = & -\delta_{i2}\delta_{j1}\left(\psi_{qq^{\prime}}^{\left(1\right)}-\psi_{qq^{\prime}}^{\left(2\right)*}\right)-\delta_{i1}\delta_{j2}\left(\psi_{qq^{\prime}}^{\left(2\right)}-\psi_{qq^{\prime}}^{\left(1\right)*}\right)\nonumber \\
 & + & \delta_{ij0}\left(\varphi_{qq^{\prime}}^{\left(1\right)}+\varphi_{qq^{\prime}}^{\left(3\right)}\right)+\delta_{ij1}\left(\varphi_{qq^{\prime}}^{\left(2\right)}-\varphi_{qq^{\prime}}^{\left(3\right)}\right)\nonumber \\
 & + & \delta_{ij2}\left(-\varphi_{qq^{\prime}}^{\left(1\right)}+\varphi_{qq^{\prime}}^{\left(4\right)}\right)+\delta_{ij3}\left(-\varphi_{qq^{\prime}}^{\left(2\right)}-\varphi_{qq^{\prime}}^{\left(4\right)}\right),\label{eq:Phi_anderson}\\
\varphi_{qq^{\prime}}^{\left(m\right)} & = & 2i\Im\sum_{k}\mathrm{Tr}_{B}\left\{ \rho_{B}\left\langle q^{\prime}\right|A_{k}^{\left(m\right)}\left|q\right\rangle \right\} ,\label{eq:anderson_phi}\\
\psi_{qq^{\prime}}^{\left(m\right)} & = & -2\sum_{k}\mathrm{Tr}_{B}\left\{ \rho_{B}\left\langle q^{\prime}\right|B_{k}^{\left(m\right)}\left|q\right\rangle \right\} ,\label{eq:anderson_psi}
\end{eqnarray}
where
$A_{k}^{\left(1\right)}=t_{1k}d_{1}d_{2}d_{2}^{\dagger}a_{1k}^{\dagger}$,
$A_{k}^{\left(2\right)}=t_{1k}d_{1}d_{2}^{\dagger}d_{2}a_{1k}^{\dagger}$,
$A_{k}^{\left(3\right)}=t_{2k}d_{1}d_{1}^{\dagger}d_{2}a_{2k}^{\dagger}$,
$A_{k}^{\left(4\right)}=t_{2k}d_{1}^{\dagger}d_{1}d_{2}a_{2k}^{\dagger}$,
$B_{k}^{\left(1\right)}=t_{1k}d_{2}a_{1k}^{\dagger}$ and
$B_{k}^{\left(2\right)}=t_{2k}d_{1}a_{2k}^{\dagger}$.  All the
quantities in \eqref{eq:Phi_anderson}--\eqref{eq:anderson_psi} are implicitly
time-dependent and can be evaluated directly with RT-PIMC~\cite{Cohen-un}.

\begin{figure}
\includegraphics[width=1\columnwidth]{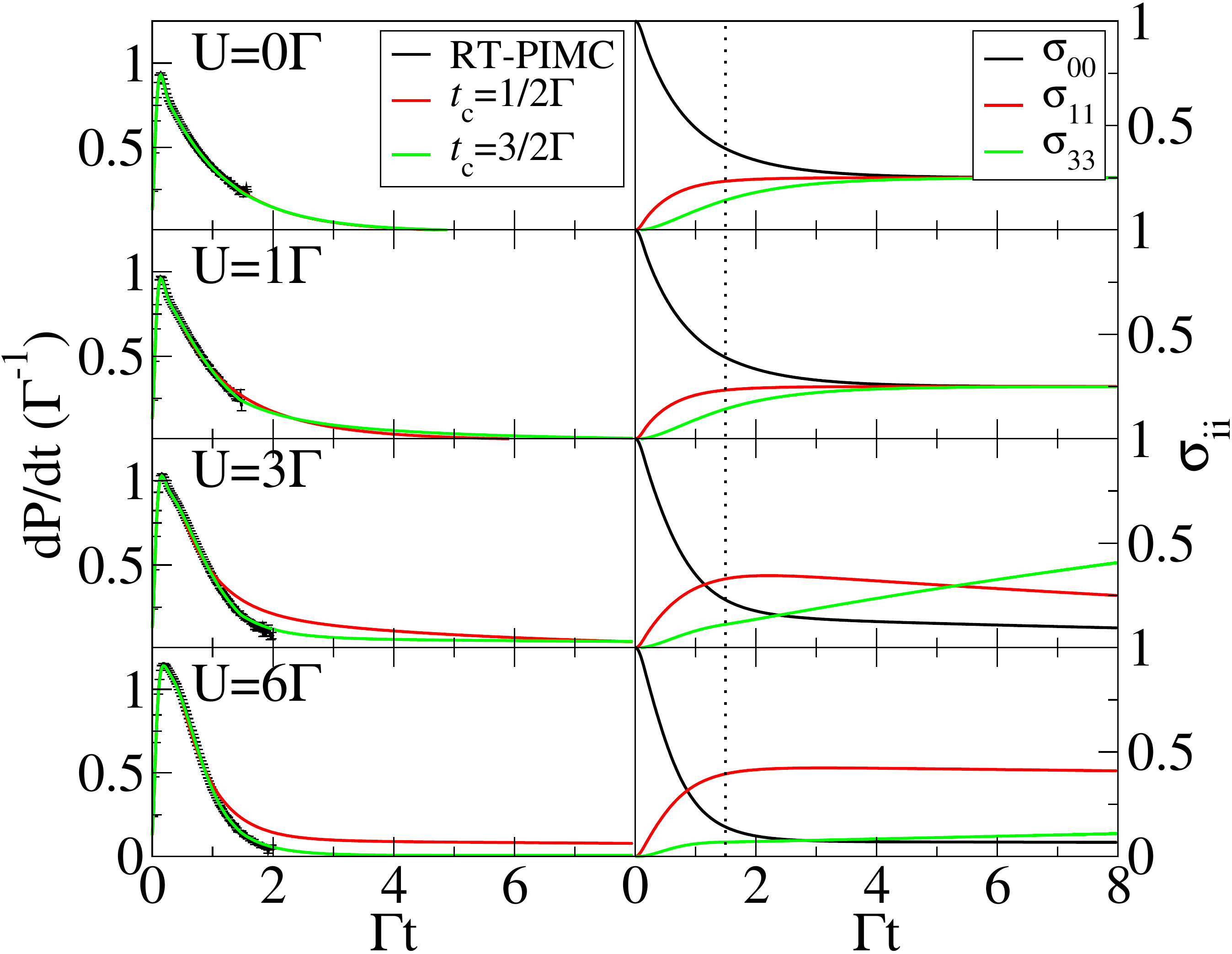}
\caption{Total population derivative from direct RT-PIMC data compared
  with the results of the memory-kernel formalism (left panels) and
  predicted dot populations (right panels) for an initially unoccupied
  dot for several values of the interaction energy $U$. Remaining
  parameters are the same as in Fig.~\ref{fig:memory-kernel}. The
  cutoff time is $\frac{1.5}{\Gamma}$, shown on the right panels as a
  vertical dotted line.}
\label{fig:populations}
\end{figure}

One can draw a few analytical conclusions directly from the block
structure of $\Phi$. First, while both the $A_{k}^{\left(m\right)}$
and $B_{k}^{\left(m\right)}$ operators introduced above conserve the
total particle number, only the $A_{k}^{\left(m\right)}$ conserve the
particle number for \emph{each} spin. Therefore, the
$\psi_{qq^{\prime}}^{\left(m\right)}$ are nonzero only if
$q,q^{\prime}=1,2$ or $2,1$ (since states $1$ and $2$ are the only dot
states which have the same total occupation, but differ in per-spin
occupation), while the $\varphi_{qq^{\prime}}^{\left(m\right)}$ can be
nonzero only when $q=q^{\prime}$. From \eqref{eq:Phi_anderson} we can
then immediately see that the diagonal density matrix elements are
coupled only to each other, with the two singly occupied off-diagonal
elements $\left|1\right\rangle \left\langle 2\right|$ and
$\left|2\right\rangle \left\langle 1\right|$ forming a second closed
block. If our interest is limited to the state populations only the
$16$ (instead of $64$) functions $\varphi_{qq}^{\left(m\right)}$ need
be evaluated within RT-PIMC.

In Fig.~\ref{fig:memory-kernel}, we plot the nonzero elements of the
memory kernel for different values of $U$. Due to the block structure
and the symmetric choice of parameters, only seven distinct nonzero
elements exist (two for $U=0$). To make the parametrization definite
within the simulations, we assume lead coupling densities of the form
$\Gamma_{i}\left(E\right)\equiv2\pi\sum_{k}\left|t_{k}\right|^{2}\delta\left(E-\varepsilon_{ik}\right)=\frac{\Gamma/2}{\left(1+e^{\nu\left(E-\varepsilon_{c}\right)}\right)\left(1+e^{-\nu\left(E+\varepsilon_{c}\right)}\right)}$,
where $\varepsilon_{c}$ is the band cutoff energy and $\nu$ is the
inverse of the cutoff width. $\Gamma$ is the maximum value attainable
by $\Gamma\left(E\right)=\sum_{i}\Gamma_{i}\left(E\right)$, as well as
its wide band limit, and will be the unit of energy throughout the
following text. We also concentrate on $\varepsilon_{i}=-\frac{U}{2}$,
known as the symmetric case of the Anderson model, yet the formalism
is general and this is certainly not a requirement The temperature
$\beta$ and the chemical potentials $\mu_{L}$ and $\mu_{R}$ enter the
calculation through $\rho_{B}$ at time zero for which we assume the
proper grand canonical distribution.

In the noninteracting case (top panel of
Fig.~\ref{fig:memory-kernel}), a rapid decay to zero is observed.
Despite the relatively broad and soft-edged band chosen here, the
decay occurs over a timescale smaller than the inverse coupling but
comparable to it. We can learn from this that any approximation based
on short memory should be expected to fail unless it can allow for
memories of at least this length, meaning for instance that Markovian
approaches to the problem cannot be expected to succeed in general.
As $U$ is increased, it becomes clear that the interaction---despite
breaking most of the symmetries between the various elements---does
not significantly affect memory decay on the first timescale.
However, even at very small interaction energies, a second, longer
timescale develops: at this timescale, a small part of each memory
kernel element decays more slowly.

The formalism becomes extremely interesting if having the memory as an
input only up to some finite cutoff time $t_{c}$---at which the
system's dynamics have not yet died out---allows accurate predictions
at far longer times. This will occur if the memory function has
essentially gone to zero by this time, such that it can be safely
truncated. In the left panels of Fig.~\ref{fig:populations} we plot
the time derivative of the total population ($dP/dt$) and show that
for certain noninteracting parameters this does indeed happen: once
$\Gamma t_{c}\gtrsim\frac{1}{3}$ dynamics at times over an order of
magnitude greater than those of the memory kernel are reliably
reproduced, and the exact steady-state result for all diagonal
elements of $\sigma$ is obtained to within the numerical errors and
shown on the top right panel of Fig.~\ref{fig:populations}. However,
as might be expected from the analysis of
Fig.~\ref{fig:memory-kernel}, in the strongly correlated cases a short
cutoff time results in incorrect populations when propagated for much
longer times than $t_{c}$, since truncating the memory at that point
has not yet become physically reasonable. For the cases of $U=\Gamma$
and $U=6\Gamma$ the qualitative physics is captured correctly within
$t_{c}=3/2\Gamma$ in that depopulation of the zero-electron level is
accelerated at short times; for $U=6\Gamma$ one also observes that at
longer times the one-electron levels draw most of the population while
the more energetic zero- and two-electron levels are suppressed.
However, for $U=3\Gamma$, the behavior predicted by the truncated
memory function is clearly wrong, consistent with the existence of
long time tails in the memory.

\begin{figure}
\includegraphics[width=1\columnwidth]{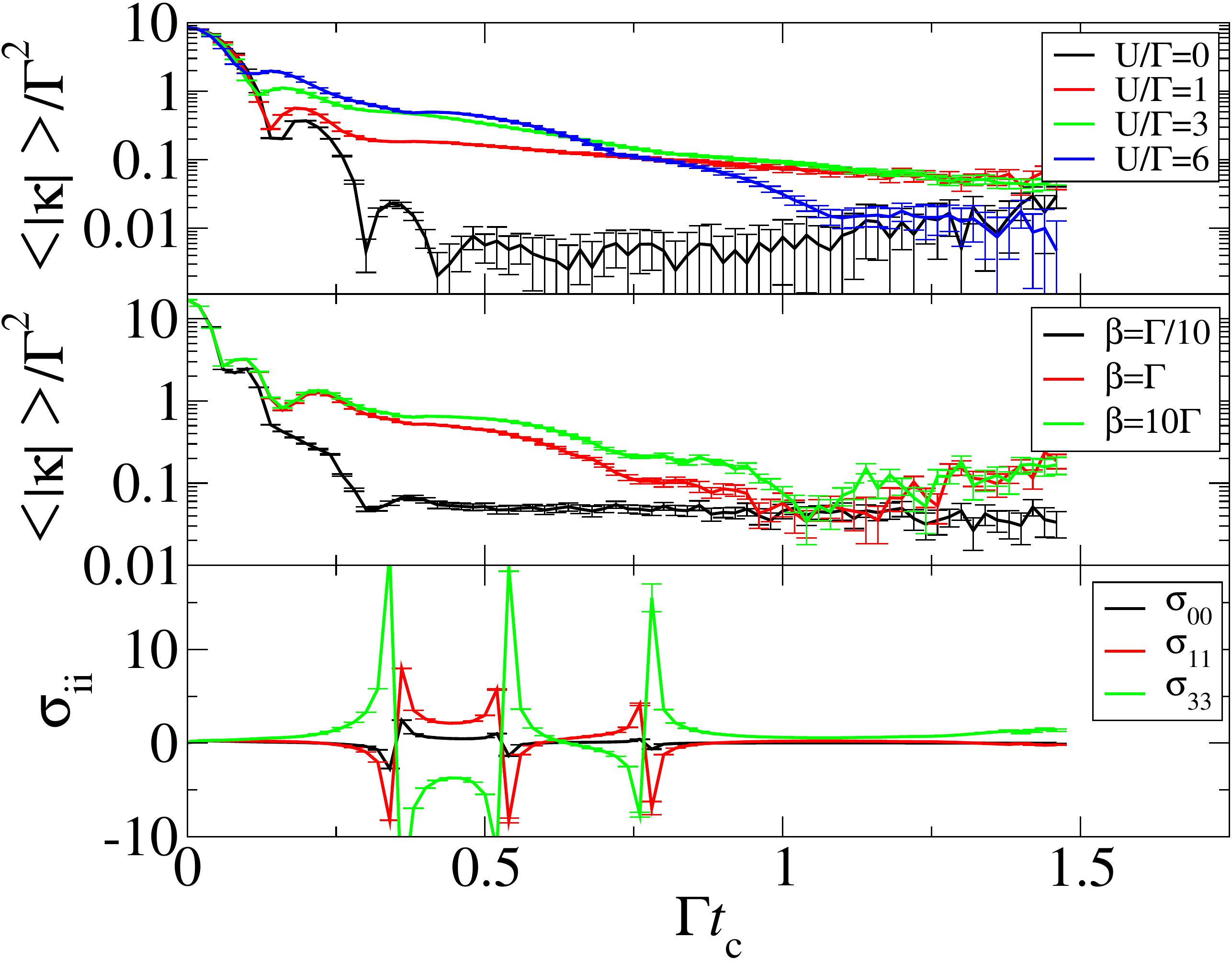}
\caption{The average absolute value of memory kernel elements at
  $\frac{\mu_{L}}{\Gamma}=-\frac{\mu_{R}}{\Gamma}=\frac{1}{2}$,
  $\frac{\varepsilon_{c}}{\Gamma}=20$, $\beta\Gamma=1$ and different
  values of the interaction energy $U$ (top panel); the same at
  $\mu_{L}=\mu_{R}=0$, $\frac{\varepsilon_{c}}{\Gamma}=40$,
  $\frac{U}{\Gamma}=6$ and different temperatures $\beta$ (middle
  panel); and the predicted steady state values of the diagonal
  density matrix elements at
  $\frac{\mu_{L}}{\Gamma}=-\frac{\mu_{R}}{\Gamma}=\frac{1}{2}$,
  $\frac{\varepsilon_{c}}{\Gamma}=20$, $\beta\Gamma=1$ and
  $\frac{U}{\Gamma}=6$.
\label{fig:kappa_size_and_steady_state}}
\end{figure}

The top two panels of Fig.~\ref{fig:kappa_size_and_steady_state}
explore the decay of the memory kernel elements more clearly by
plotting the average absolute value of the memory kernel elements on a
logarithmic scale, for various values of $U$, $\varepsilon_{c}$ and
$\beta$.  Notably, while only the noninteracting case appears to have
a memory kernel which goes to within the numerical errors of zero
within the simulation timescale shown, strong interaction actually
appears to reduce the memory lifetime when compared with intermediate
values.  This relates to the fact the $U=3\Gamma$ problem is "harder"
in this sense than the strongly interacting $U=6\Gamma$ problem, as
discussed above. While increasing either the band width or the
temperature appears to affect the short-term behavior, reducing the
shorter memory timescale, the longer timescale appears largely
unaffected by the variation of these parameters. This observation
seems to be consistent with the hypothesis that the timescale of the
tail's memory decay is related to the inverse Kondo
temperature. However, we find that a large bias voltage does not
affect markedly the longer timescale, despite supposedly destroying
the Kondo correlations. 

In addition to the time dependence, one can obtain the steady state
result directly from the stationary state equation
\begin{equation}
\left[\mathcal{L}_{S}-\frac{i}{\hbar}\hat{\kappa}\left(z\rightarrow
  i0\right)\right]\sigma\left(t\rightarrow\infty\right)=\lim_{z\rightarrow0}z\hat{\vartheta}\left(z\right)
\label{eq:stationary_state_equation}
\end{equation}
with the added condition that $\mathrm{Tr}\sigma=1$. This means that
for an initially uncorrelated system, a unique steady state exists if
and only if the supermatrix $\mathrm{Tr}_{S}\left\{
\left(\left|i\right\rangle \left\langle
j\right|\right)^{\dagger}\left[\mathcal{L}_{S}-\frac{i}{\hbar}\hat{\kappa}\left(z\rightarrow
  i0\right)\right]\left|k\right\rangle \left\langle l\right|\right\} $
has a degeneracy of exactly one. In the bottom panel of
Fig.~\ref{fig:kappa_size_and_steady_state}, the steady state values
obtained from this formula for parameters close to the Kondo regime
are plotted against the cutoff time $t_{c}$. While the trace of the
density matrix is conserved, physically impossible results appear at
intermediate cutoff times and convergence is not yet achieved. The
long tails of the memory kernel elements are therefore crucially
important for the correct prediction of both dynamical and
steady-state properties.

In conclusion, we have developed a numerically exact method for the
formulation and solution of the reduced dynamics of quantum impurity
models, and applied it to the nonequilibrium Anderson model. It is
clear from our results that the physics of even a noninteracting
electronic junction cannot be fully captured within a Markovian
picture, and that on-site interaction results in deeply non-Markovian
physics even at relatively large band widths, bias voltages and
temperatures.  We show that the long memory tails induced by the
Hubbard term affect both the dynamics and the steady-state, despite
their relative smallness.  In the computational sense, the proposed
method is extremely useful in extending the applicability of RT-PIMC
to long time scales and steady state when the memory goes to zero
within the simulation time scale, but the dynamics do not.

The authors would like to thank Abe Nitzan and David Reichman for
useful discussion. GC is grateful to the Azrieli Foundation for the
award of an Azrieli Fellowship. ER thanks the Miller Institute for
Basic Research in Science at UC Berkeley for partial financial support
via a Visiting Miller Professorship. This work was supported by the
US--Israel Binational Science Foundation and by the FP7 Marie Curie
IOF project HJSC.

\end{document}